\documentclass[aps,prl,twocolumn,groupedaddress,showpacs]{revtex4}
\usepackage{amsbsy,amssymb}
\usepackage{graphicx}
\usepackage{eucal}
\newcommand{\hana}[1]{{\mathcal{#1}}}
\newcommand{\dd}{{\rm d}}
\newcommand{\DD}{{\rm D}}
\newcommand{\eexp}{{\rm e}}

\begin{document}

\title{Species Abundance Patterns in Complex Evolutionary Dynamics}

\author{Kei Tokita}
\email[]{tokita@cmc.osaka-u.ac.jp}
\homepage[]{http://www.cp.cmc.osaka-u.ac.jp/~tokita}
\thanks{Permanent address: Large-Scale Computational Science Division,
Cybermedia Center, Osaka University, 1-32 Machikaneyama-cho, Toyonaka,
Osaka 560-0043, Japan}
\affiliation{Program for Evolutionary Dynamics, Harvard University, One Brattle Square, Cambridge, MA  02138, USA}

\date{\today}

\begin{abstract}
An analytic theory of species abundance patterns (SAPs) in biological
networks is presented. The theory is based on multispecies replicator dynamics
equivalent to the Lotka-Volterra equation, with diverse
interspecies interactions. Various SAPs observed in nature are derived
from a single parameter. The abundance distribution is formed like a widely observed left-skewed lognormal
distribution. As the model has a general form, the result can be
applied to similar patterns in other complex biological networks,
e.g. gene expression.
\end{abstract}

\pacs{87.23.-n,75.10.Nr,87.10.+e,87.90.+y}

\maketitle

If we investigate the number and populations of species in an ecosystem,
we can observe universal characteristic patterns in that ecosystem. How
to clarify the mechanisms underlying those species abundance patterns
(SAPs) has been one of the 'unanswered questions in ecology in the last
century \cite{May_1999}' even though the knowledge obtained from it
would affect vast areas of nature conservation. Various models have been
applied to ecosystem communities where species compete for niches on a
trophic level
\cite{Motomura_1932,Corbet_Fisher_Williams_1943,MacArthur_1960,Preston_1962b,Whittaker_1970,Bazzaz_1975,May_1975,Sugihara_1980,Nee_Harvey_May_1991,Tokeshi_1999,Hubbel_2001,Hall_etal_2002,McGill_2003,Volkov_etal_2003,Pigolotti_etal_2004},
but these models have left the more complex systems a mystery. Such
systems occur on multiple trophic levels and include various types of
interspecies interactions, such as prey-predator relationships,
mutualism, competition, and detritus food chains. Although SAPs are
observed universally in nature, their essential parameters have not
been fully clarified.

I consider, then, a widely adopted model of biological networks
represented by the so-called  $N$-species replicator equation (RE) \cite{Hofbauer_Sigmund_1998};
\begin{equation}
\frac{\dd x_i}{\dd t} = x_i \left(
                             \sum_j^N J_{ij}x_j -
                             \frac{1}{N}\sum_{j,k}^N J_{jk}x_jx_k
                            \right)_,
\label{RE}
\end{equation}
to calculate the abundance $x_i(t)(\in [0, N])$ of species
$i(=1,2\ldots, N)$.  Here we assume that $(J_{ij})$ is a
time-independent random symmetric $(J_{ij}=J_{ji})$ matrix whose
elements have a normal distribution with mean $m (>0)$ and variance
$\tilde{J}^2/N$ as
\begin{equation}
P(J_{ij})=\sqrt{\frac{N}{2\pi\tilde{J}^2}}\exp\left[
               - \left(
                   \frac{N}{2\tilde{J}^2}
                \right)\left(J_{ij}-m\right)^2
                                              \right]_.
\end{equation}
Self-interactions are all set to a negative constant as $J_{ii}=-u
(<0)$.  Note that the essential parameter is unique as $p\equiv
(u+m)/\tilde{J}$ because the transformation of the interaction
$K_{ij}\equiv (J_{ij}-m)/\tilde{J}$ does not change the trajectory of
the dynamics (\ref{RE}).  Although ecologists do not generally believe
in the randomness of interspecies interactions in nature, the discipline
has been affected by the random interaction model \cite{May_1972} as a
prototype of complex systems.

The RE appears in various fields \cite{Hofbauer_Sigmund_1998}. In
sociobiology, it is a game dynamical equation for the evolution of
behavioral phenotypes; in macromolecular evolution, it is the basis of
autocatalytic reaction networks ({\it hypercycles}); and in population
genetics it is the continuous-time selection equation in the symmetric
$(J_{ij}=J_{ji} )$ case. 
The symmetric RE also corresponds to a classical
model of competitive community for resources\cite{MacArthur_Levins_1967}.

 Particularly in the context of ecology, the
$N-1$ species Lotka-Volterra (LV) equation
\begin{equation} 
 \frac{\dd y_i}{\dd t}=y_i\left(
                       r_i - \sum_j^{N-1}b_{ij}y_j
                          \right)
\end{equation}
is equivalent to the $N$ species RE \cite{Hofbauer_Sigmund_1998}.
That is, the abundance $y_i$ and the parameters
in the corresponding LV are described by those in
the present RE model as,
\begin{eqnarray}
&&y_i = x_i/x_M \quad (i=1,2,\ldots ,N)\\
&&r_i = J_{iM} - J_{MM} = J_{iM} + u\label{ri}\\
&&b_{ij} = J_{ij}-J_{Mj}\label{bij}
\end{eqnarray}
where the 'resource' species $M\, (y_M=1)$ can be arbitrarily chosen
from $N$ species in the RE. The ecological interspecies interactions
$(b_{ij})(i\neq j)$ have a normal distribution with mean $0$ and
variance $2\tilde{J}^2/N$ from Eq.~(\ref{bij}), and they are no longer
symmetric ($b_{ij}\neq b_{ji}$). The present model therefore describes
an ecological community with complex prey-predator interactions
$((b_{ij}, b_{ji})\to (+,-)\,\mbox{or}\, (-,+))$, mutualism $(+,+)$ and
competition $(-,-)$. Moreover, a community can have a 'loop' (detritus)
food chain $((b_{ij}, b_{ji})\to (+,-), (b_{jk}, b_{kj})\to (+,-),
(b_{ki}, b_{ik})\to (+,-))$.  The intraspecific interaction $b_{ii}$
turns out to be related to the intrinsic growth rate $r_i$ as
$b_{ii}=J_{ii}-J_{Mi}=-u-J_{iM}=-r_i$ and is therefore competitive
$(b_{ii}<0)$ for producers $(r_i>0)$ or mutualistic $(b_{ii}>0)$ for
consumers $(r_i<0)$.

By Eq.~(\ref{ri}), the intrinsic growth rates also have a normal
distribution with mean $u+m$ and variance $\tilde{J}^2/N$. The
probability at which $r_i$ is positive--that is, that the
$i$-th species is a producer--is therefore given by the
error function,
\begin{equation} 
 \mbox{Prob}(r_i>0)=\int_{-p\sqrt{N/2}}^\infty 
   \frac{\dd t}{\sqrt{\pi}}\exp\left(-t^2\right)_.
\end{equation}
Consequently, the parameter $p$ can be termed as the 'productivity' of
a community because the larger the $p$, the greater the number of
producers. The parameter $p$ is also connected to the maturity of an
ecosystem because $m$ increases in time in an evolutionary model
\cite{Tokita_Yasutomi_2003}.

The symmetry $(J_{ij}=J_{ji})$ makes the average fitness
$\bar{f}\equiv\sum_{j,k}^NJ_{jk}x_jx_k$ (the second term of the
r.h.s. of Eq.~(\ref{RE})) a Lyapunov
function \cite{Hofbauer_Sigmund_1998}, which is a nondecreasing function
of time in dynamics (\ref{RE}). Therefore, every initial state
converges to a local maximum of $\bar{f}$ as $t\to\infty$. Interpreting
$\hana{H}\equiv -\frac{1}{2}\bar{f}$ as an energy function, we can study
macroscopic functions like free energy at such a maximum by using
the technique of statistical mechanics of random systems \cite{Mezard_etal_1987,Diederich_Opper_1989,Biscari_Parisi_1995,de_Oliveira_Fontanari_2000,de_Oliveira_Fontanari_2001,de_Oliveira_Fontanari_2002}.

Information on equilibrium states of dynamics (\ref{RE}) at
$t\to\infty$ is derived from
the zero-temperature limit of free energy density,
\begin{eqnarray}
f&\equiv &
 -\lim_{\beta\to\infty}\lim_{N\to\infty}\frac{1}{N\beta}\left[\ln
                                                        Z\right]_J\label{f}\\
Z&\equiv &
 \int_0^\infty\left(\prod_i^N\dd x_i\right)\delta(N-\sum_k^N
               x_k)\eexp^{-\beta\hana{H}}\label{Z}
\end{eqnarray}
 where $\left[\ldots\right]_J\equiv\int_{-\infty}^\infty\dd J_{ij}P(J_{ij})(\ldots)$
denotes the 'sample average \cite{Mezard_etal_1987}' over random
interactions. The Dirac delta function in
Eq.~(\ref{Z}) reflects the conservation of total abundance
$\sum_i^Nx_i(t)=N$ satisfied at any $t$ in
Eq.~(\ref{RE}). The calculation of Eq.~(\ref{f}) is similar to calculations in the
previous
works \cite{Diederich_Opper_1989,Biscari_Parisi_1995,de_Oliveira_Fontanari_2000,de_Oliveira_Fontanari_2001,de_Oliveira_Fontanari_2002}
and yields mean field equations for the order parameters $q$ and $v$ as
\begin{eqnarray}
&&p-v =\sqrt{q}
\int_{-\Delta}^\infty
(z+\Delta) \DD z\label{mfeq1}\\
&&(p-v)^2 = \int_{-\Delta}^\infty
(z+\Delta)^2 \DD z\label{mfeq2}
\end{eqnarray}
where $\Delta\equiv\sqrt{q}(p-2v)$ and $\DD z\equiv \dd
 z\exp(-z^2/2)/\sqrt{2\pi}$.
The resulting equations turn out to be formally the same
as the case where $m=0$ and $\tilde{J}=1$ \cite{Diederich_Opper_1989}. 
For each value of $p$, Eqs.~(\ref{mfeq1}) and (\ref{mfeq2}) are solved
numerically.  

Among macroscopic functions calculated in the present
framework, the most significant for a theory of SAPs is the {\it survival
function}
$\alpha_p(x)\equiv\frac{1}{N}\sum_i^N\theta (x_i-x)$, 
the proportion of species whose abundance is larger than $x$,
where $\theta(z)(=1 (z>0); 0 (z\leq 0))$ is the step function. 
Similar to the fashion in which free energy
was calculated, the survival function $\alpha_p(x)$ is analytically calculated and
represented by the order parameters as 
\begin{eqnarray}
\alpha_p(x)&\equiv&\lim_{\beta\to\infty}\lim_{N\to\infty}\left[
                \int_0^\infty\left(\prod_j^N\dd x_j\right)\right.\nonumber\\
&&\left.\,\times\;
                \theta (x_i-x)\delta\left(N-\sum_k^Nx_k\right)
                \frac{\exp(-\beta\hana{H})}{Z}
                                                    \right]_J\nonumber\\
&&\!\!\!\!\!\!\!\!\!\!\!\!=\alpha_p(0)\tilde{\theta}(x)
   -\int_{-\Delta}^\infty
    \tilde{\theta}\left(
    x - \frac{\sqrt{q}(z+\Delta)}{p-v}
                \right)\DD z
\end{eqnarray}
where the definition of the step function above is given by
$\tilde{\theta}(z)=1 (z\geq 0); 0 (z<0))$. 

\begin{figure}
\includegraphics[width=8.6cm]{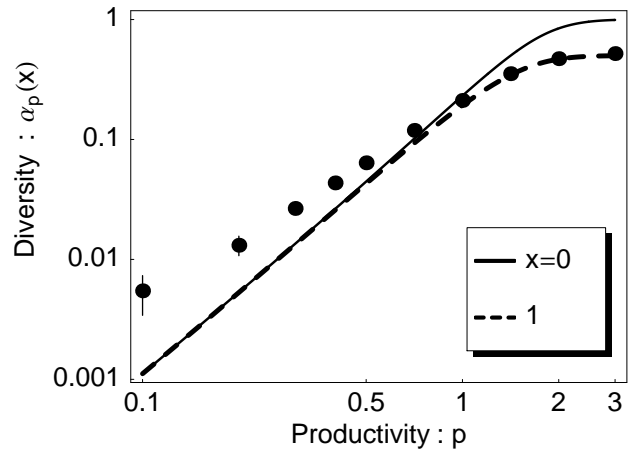} \caption{\label{Figure1}
Diversity $\alpha_p(x=0,1)$ as a function of $p$ of log-log
scales. Black circles show numerical solutions of $\alpha_p(1)$
averaged over $50$ samples of $(J_{ij})$ for Eq.~(\ref{RE}) with
$N=2048$ and $p=0.1, 0.2, 0.3, 0.4, 0.5, \sqrt{2}/2, 1, \sqrt{2}, 2,
3$. Error bars indicate the maximum and minimum values found in the
samples.}
\end{figure}

The resulting function $\alpha_p(0)\equiv v(p-v)$ and $\alpha_p(1)$ of
$p$ can be termed 'diversity', i.e., the proportion of nonextinct
species and that of the species with abundance larger than unity,
respectively, as depicted in Fig.~\ref{Figure1}. This demonstrates a
typical positive correlation between productivity and diversity
\cite{Waide_etal_1999}. Numerical results for $\alpha_p(1)$ are also
depicted in Fig.~\ref{Figure1} for comparison.  We see good agreement
between the analytical and the numerical results for $p\gtrsim 1$, while
some deviations appear for small values of $p$. This small-value
deviation is attributable to the occurrence of replica symmetry breaking
(RSB)\cite{Mezard_etal_1987,Biscari_Parisi_1995} for $p<\sqrt{2}$, which
yields a number of metastable states of Eq.~(\ref{f}), and the
replicator dynamics (\ref{RE}) essentially converges to not only a
ground state of (\ref{f}) but also to the metastable states. Since the
energy $\hana{H}$ and the diversity are both nonincreasing functions of
time in dynamics (\ref{RE}), the mean-field results here give a lower
minimum of diversity. Interestingly, the metastable states enhance the
diversity. The analysis of RSB is expected to improve the quantitative
agreement \cite{Biscari_Parisi_1995}.

\begin{figure}
\includegraphics[width=8.6cm]{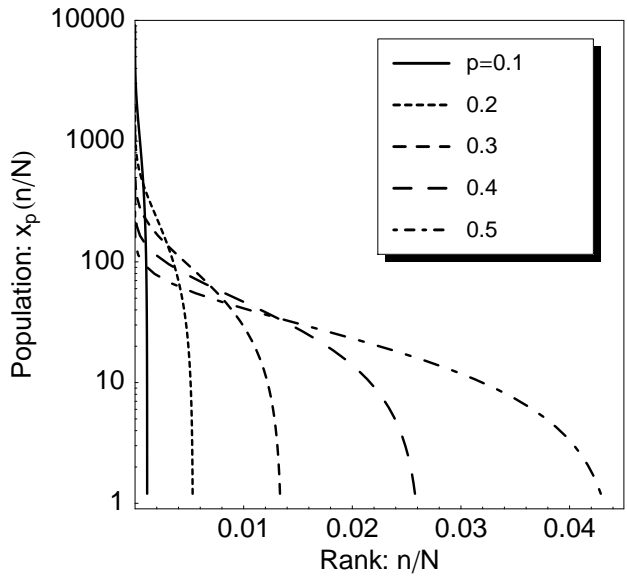}
\caption{\label{Figure2} Rank-abundance relations as a function of productivity $p$ on normal-log scales.}
\end{figure}

Note that $\alpha_p(x)$ is also represented as a
function of species rank $n$: 
\begin{equation}
\alpha_p(x)=\frac{n}{N}\quad\mbox{for}\quad x\in [x^{(n+1)}, x^{(n)})_,
\end{equation}
$(n=1,2,\ldots , S\le N)$ if the species abundance is ranked in
descending order, as in $x^{(1)}\ge x^{(2)}\ge\cdots\ge
x^{(n)}\ge\cdots\ge x^{(S)}>0$. As the function $\alpha_p(x)$ is a
nonincreasing monotonic function, the species abundance relation, i.e., the abundance $x^{(n)}$ as a
function of a rank $n$, is given by the inverse function of
$\alpha_p(x)$ as 
$x^{(n)}=x_p(n/N)=\alpha_p^{(-1)}(x)$,
depicted in Fig.~\ref{Figure2}
for some values of $p$. We observe two typical SAPs in different regions
\cite{Hubbel_2001} and with different species compositions
\cite{Whittaker_1970}: one is a straight line like the geometric series
\cite{Motomura_1932} for a small value of $p$, and the other consists of
sigmoid curves on a logarithmic vertical axis for some range of
$p$. This latter SAP denotes a lognormal-like abundance
distribution. Remarkably, the transition of the SAPs from low $p$ to
high is identical to the observed transition from low- to
high-productivity areas; that is, from a species-poor area such as an
alpine or polar region to a species-rich tropical rain forest
\cite{Hubbel_2001}. The transition also corresponds to the secular
variation of SAPs observed in abandoned cultivated land
\cite{Bazzaz_1975}. This supports the contention that $p\; (\mbox{or} \, m)$ is a maturity parameter,
as is suggested by an evolutionary model \cite{Tokita_Yasutomi_2003}.

\begin{figure}
\includegraphics[width=8.6cm]{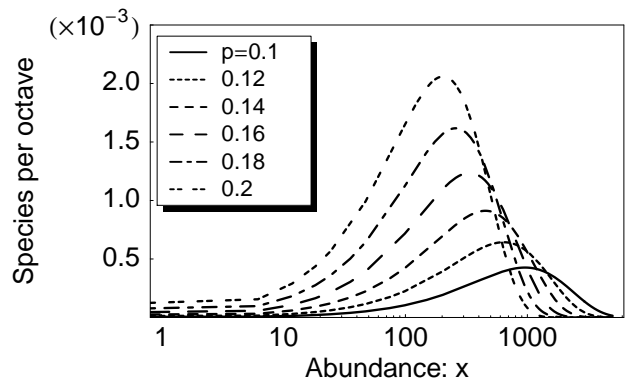} \caption{\label{Figure3}
Abundance distribution per 'natural' octave $\ln(x)$. Functions
$F_p(x)x(\dd (\ln(x))=F_p(x)\dd x)$ for some values of $p$ are depicted,
whereas Preston originally defined octaves as logarithms to base 2
\cite{Preston_1962b}.}
\end{figure}

The abundance distribution is also derived from
the survival function. As $C_p(x)\equiv 1-\alpha_p(x)$ is a cumulative distribution function of
abundance, the abundance distribution is given by the derivative
$F_p(x)\equiv\dd C_p(x)/\dd x$ and
\begin{eqnarray} 
F_p(x)&=&\frac{p-v}{\sqrt{2\pi q}}\exp\left\{
    -\frac{(p-v)^2}{2q}\left(
      x - \frac{q(p-2v)}{p-v}
                       \right)^2
                                \right\}\nonumber\\
  &&\, +\,\, C_p(0)\delta(x)
\end{eqnarray}
where the second term denotes the rate of species extinction. The first
term is a normal distribution but not a lognormal
distribution. Nevertheless, the curves in Fig.~\ref{Figure2} demonstrate
a typical sigmoid pattern on a logarithmic vertical axis. This pattern
indicates the coexistence of very abundant species with rare ones. This
multiscale of abundance is intuitively understood by a divergent
behavior of the variance $\sigma^2\equiv q/(p-v)^2$ of $F_p(x)$ for
small $p$ because $q\to\infty$ and $v\to 0$ for $p\to 0$. Moreover, the
mode of $F_p(x)$ per 'natural' octave \cite{Preston_1962b} $\ln (x)$
is always a positive value (as shown in Fig.~\ref{Figure3}) at
$x^*=\frac{\sigma}{2}(\Delta + \sqrt{\Delta^2+4})>0$, which denotes a
unimodal distribution. Indeed, the mode diverges as
\begin{equation}
x^*\to\frac{\sigma}{|\Delta |}=\frac{1}{(p-v)|p-2v|}\to\infty
\end{equation}
for $p\to 0$. As a result, the abundance
distribution is a truncated normal distribution with a large variance $\sigma^2\to\infty$
and a negatively divergent mean $\mu\equiv\frac{q(p-2v)}{p-v}\to -\infty$
satisfying $\frac{\sigma}{\mu}=\frac{1}{\Delta}\to 0$ for $p\to 0$. This is why the
abundance distribution per octave looks like a left-skewed lognormal
distribution \cite{Nee_Harvey_May_1991} in Fig.~\ref{Figure3}.  

Moreover, we derive an analytical expression for the population
$x_{max}$ of the most abundant species, the position of the individual
curve mode $x_N$, and therefore the ratio of their logarithm defined
as $\gamma\equiv\log(x_N)/\log(x_{max})$. According to the {\it
canonical hypothesis} \cite{Preston_1962b,May_1975}, the parameter
$\gamma$ takes a value near unity in various real communities. To
check the validity of the canonical hypothesis, we first need to
evaluate an expected value of the most abundant species
$x_{max}$. From the definition of $x_{max}$, that is
$NF_p(x_{max})=1$, and the conservation of the total abundance
$\int_0^\infty F_p(x)x\dd x=N$, which is equivalent to
$\sum_i^Nx_i=N$, we obtain
\begin{equation}
x_{max}=\frac{q(p-v)+\sigma\sqrt{2\ln\left(\frac{\sigma(1-\alpha_p(0))+\Delta\alpha_p(0)}{\sqrt{2\pi}}\right)}}{p-v}_.
\end{equation}
On the other hand, the mode of the individual curve $F_p(x)x$ per
octave is given by $x_N=\frac{\sigma}{2}\left(\Delta
+\sqrt{\Delta^2+8}\right)$, and finally, the parameter
$\gamma\equiv\log(x_N)/\log(x_{max})$ is evaluated by substituting the
values of the order parameters $q$ and $v$ for each value of $p$.  In
the present model, $\gamma$ is a monotonically increasing function of
$p$ and $0.96<\gamma <1.04$ for $0.1<p<0.6$, denoting that the
canonical hypothesis is supported in the range of $p$ giving the
typical SAPs in Fig.~\ref{Figure3}. Although the canonical hypothesis
was demonstrated to be merely a mathematical consequence of lognormal
distribution \cite{May_1975} rather than anything biological, it is
noteworthy that the lognormal-like abundance distribution with
$\gamma\simeq 1$ derives from basic ecological dynamics. This still
suggests a biological foundation for the hypothesis in a large complex
ecosystem, in the same way that a biological foundation was indicated
for the theory of a local competitive community \cite{Sugihara_1980}.

In the present model, all species coexist only in the limit
$p\to\infty$, that is, in the trivial cases in which interspecies
interactions are negligible ($\tilde{J}\ll u$ ) or homogeneous
($\tilde{J}\to 0$), thereby giving $\alpha_\infty (x) = \theta(1-x)$,
$x^{(n)}=x_\infty (n/N)=1$ for all $n$ and $F_\infty(x)=\delta(x-1)$.

The present theory seeks to capture the influence of productivity on the
SAPs under the assumption that all species interact randomly;
nevertheless, this assumption itself is never justified because it
ignores a biological correlation between interactions produced by
evolution. However, note that the randomness is assumed only for an
initial state with $N$ species in Eq.~(\ref{RE}). Actually, the
simulation reveals the resulting interactions of nonextinct species
to be nonrandom: every sample for
$p\in[0.1,3]$ in Fig.~\ref{Figure1} evolves to only flora, $\forall i\,\,
r_i>0$. Moreover, by ordering species as $r_i>r_j$ for any $i<j$, we
observe a hierarchy: there are only three types of interactions, that is,
mutualism $(b_{ij},b_{ji})=(+,+)$, competition $(-,-)$ and exploitation
of $i$ on $j$ as $(+,-)$, but no reverse $(-,+)$. This suggests the
applicability of the present model to a plant community.

It has been demonstrated that empirically supported patterns
are derived from a single parameter of general population dynamics.
This not only suggests the
importance of globally coupled biological interactions in a large
assemblage but also provides a unified viewpoint on mechanisms of
similar patterns observed in other biological networks with
complex interactions; for example, a lognormal abundance distribution
of a protein in
cells \cite{Blake_etal_2003,Kaneko_PRE_2003,Sato_etal_2003}, which is
revealed by gene expression networks.

\begin{acknowledgments}
The author thanks R. Frankham, Y. Iwasa, E. Matsen, R. May, M. Nowak and
J. Plotkin for their helpful comments. This work was supported by
Grants-in-Aid from MEXT, Japan.
\end{acknowledgments}


\end{document}